\documentclass{jetpl}
\twocolumn
\title{Solution of the problem of catastrophic relaxation of homogeneous
spin precession in superfluid $^3$He-B}

\rtitle{Catastrophic relaxation of homogeneous precession in
$^3$He-B}

\sodtitle{Solution of the problem of catastrophic relaxation of
homogeneous spin precession in superfluid $^3$He-B}

\author{Yu.\,M.\,Bunkov
$^{a}$\/\thanks{e-mail: yuriy.bunkov@grenoble.cnrs.fr},
V.\,S.\,L'vov$^{bc}$\/\thanks{email: Victor.Lvov@Weizmann.ac.il},
G.\,E.\,Volovik$^{cd}$\/\thanks{e-mail: volovik@boojum.hut.fi}}

\rauthor{Yu.\,M.\,Bunkov, V.\,S.\,L'vov, G.\,E.\,Volovik}

\sodauthor{Bunkov, Lvov, Volovik}

\address{$^{a}$ Centre de Recherches sur les Tr\`es Basses Temp\'eratures,
 CNRS, BP166, 38042, Grenoble, France
\\~\\$^{b}$ Department of
Chemical Physics, The Weizmann Institute of Science, Rehovot 76100,
Israel
\\~\\
$^{c}$ Low Temperature Laboratory, Helsinki University of
Technology,
P.O.Box 2200, FIN-02015, HUT, Finland\\~\\
$^{d}$ Landau Institute for Theoretical Physics RAS, Kosygina 2,
117334 Moscow, Russia}

\dates{14 May 2006}{*}

 \abstract{The quantitative analysis of the
``catastrophic relaxation" of the coherent spin precession in
$^3$He-B is presented. This phenomenon has been observed below the
temperature about 0.5 T$_c$ as an abrupt shortening of the induction
signal decay. It is explained in terms of  the decay
 instability  of homogeneous transverse NMR mode into  spin waves of the
longitudinal NMR.  Recently the cross interaction
amplitude between the two modes has been calculated by Sourovtsev
and Fomin \cite{SF} for the so-called Brinkman-Smith configuration,
i.e. for the orientation of the orbital momentum of Cooper pairs
along the magnetic field, ${\bf L}\parallel {\bf H}$. In their
treatment, the interaction is caused by the anisotropy of the speed
of the spin waves. We found that in the more general case of the
non-parallel orientation of ${\bf L}$ corresponding to the typical
conditions of experiment, the spin-orbital interaction provides the
additional interaction between the modes. By analyzing experimental
data we are able to distinguish which contribution is dominating in
different regimes.}


\begin{document}

\maketitle


\section{Introduction}

For magnetically ordered systems the instability of homogeneous
precession is a well known phenomenon. Suhl \cite{Suhl} explained
  it in terms of parametric instability of the mode of precession
with respect to excitations of pairs of spin waves satisfying the
condition of resonance:
  \begin{equation}
  n\omega_L=\omega_s({\bf k}) +\omega_s(-{\bf k})~,
\label{resonance}
  \end{equation}
where $\omega_L$ is the precession frequency and $n$ is integer (see
also  book \cite{NSW} ). Parametric instability (see, e.g.
textbook by Landay and Lifshts,~\cite{LL}) is a particular case of
the decay instability, which  is very general and well known in
physics of nonlinear wave phenomenon, that includes induced light
scattering (photon decaying into photon and phonon or spin wave)
decay of Lengmuir waves into Lengmuir and ion-sound waves in
non-isotherm plasmas, decay of capillary wave into capillary and
gravity wave on the water surface,  etc. In   quantum solids and
liquids the decay instability has been observed in
anti-ferromagnetic solid $^3$He \cite{Mizusaki}, and has been
predicted for superfluid liquid
$^3$He-A \cite{HeA} where it has been observed later \cite{HeAExp}.

 Unique feature of superfluid $^3$He-B  is  the  phenomenon
 self sustained and long-lived precession with the coherent phase
across the whole precessing domain, the so-called Homogeneously
Precessing Domain (HPD)\cite{HPD}. What is the really  amazing is a
huge precession angle up to $104^{\rm o}$ in the presence of
lower-frequency spin waves (longitudinal NMR), for which the
resonance conditions~(\ref{resonance}) are definitely satisfied.  In
general nonlinear media the life time of this kind of excitation
(with dimensionless amplitude of order unity) should be about their
period.  The main question is why the HPD is so long-living
excitation? The explanation is unique symmetry of the leading Zeeman
interaction which therefore does not contribute to interaction amplitude
of the decay processes~(\ref{resonance}) [that we denote as $V(k)$].
Moreover, even subleading spin-orbit interaction in the Brinkman-Smith
configuration (with ${\bf L}\parallel {\bf H}$) does not contribute to
the interaction amplitude $V(k)$ due to the symmetry constraint.

Nevertheless  the abrupt instability of HPD and of the
homogeneous precession in general, called the catastrophic
relaxation, has been observed in  $^3$He-B below
$\sim (0.4-0.5) T_c$ \cite{CatHPD}.  The main physical question
here is to clarify what  is the origin of contribution to
interaction amplitude  $V(k)$ that  is responsible   for the
catastrophic relaxation in general and in the experimental
conditions \cite{CatHPD} in particular. First reasonable step in
this direction was recently made in  \cite{SF} who considered simple
${\bf L}\parallel {\bf H}$ configuration and found non-vanishing
contribution to $V(k)$, originated from the dependence of the
spin-wave velocity on the direction of propagation. As we will show
below this contribution hardly can be considered as the main one in
typical experimental conditions~\cite{CatHPD}.

 In our Letter
we found  another
contribution to the decay-interaction~(\ref{resonance}) amplitude
$V(k)$   [denoted below as $ V_{_{\rm BLV}}(k)$] that allows one to
rationalize main features of observed catastrophic relaxation. Our
point is that under conditions of the experiment~\cite{CatHPD}, the
boundary conditions on the wall of container induce the texture of
the order parameter in which the orbital vector ${\bf L}$ deviates
from its symmetric orientation along the magnetic field ${\bf H}$
 in the most of the container volume. The symmetry of the
spin-orbit interaction is violated providing the additional term in
the interaction $V(k)$ between the modes, which is dominating in
typical experiments with the catastrophic relaxation
\cite{CatHPD,HP1, HP2}.

\section{General precessing states}

The homogeneous precession of magnetization in $^3$He-B has been
analyzed by Brinkman and Smith \cite{BS} and in a great detail by
Fomin \cite{FBS} for the ideal case of the symmetric orientation of
the orbital momentum ${\bf L}\parallel {\bf H}$. To discuss the real
experiments in which the orbital vector ${\bf L}$ is deflected due
to the boundary conditions and forms the texture, the computer
simulations on the basis of the full set of Leggett-Takagi equations
for spin dynamics have been employed, using the program elaborated
by Golo \cite{BG} for the one dimensional texture. The instability
of homogeneous precession was demonstrated: it was found that at the
temperatures, corresponding to the experimentally observed
catastrophic relaxation, some mode with fixed non-zero wave vector
$k$ starts to grow exponentially. It was also found \cite{B} that
the magnetic field dependence of the onset of instability is in
quantitative agreement with the field dependence of catastrophic
relaxation observed in \cite{Lee}. This demonstrates that the
catastrophic relaxation is indeed in the frame of the Leggett-Takagi
equations, but for the real understanding of this phenomenon we must
identify the principle mechanism of this calculated instability. For
that we modified the
theory \cite{SF} for
the general case of arbitrary orientation of the orbital vector
${\bf L}$.

\subsection{Symmetry of precessing states}

Let us consider the general homogeneous free precession in external
magnetic field ${\bf H}$. In liquid $^3$He the spin-orbit
(dipole-dipole) interaction is weak. If it is neglected, we can
apply the powerful Larmor theorem, according to which, in the
spin-space coordinate frame  rotating with the Larmor frequency the
effect of magnetic field on spins of the $^3$He atoms is completely
compensated. Since the magnetic field becomes irrelevant, the
symmetry group of the physical laws in the precessing frame is

\begin{equation}
  G= SO_3^{L}\times SO_3^{S} ~~,
\end{equation}

where $SO_3^{L}$ is the group of orbital rotations in the laboratory
frame; and $SO_3^{S}$ is the group of spin rotations in the rotating
frame whose elements ${\bf \tilde g(t)}$ are constructed from the
elements ${\bf g}$ of conventional spin rotations in the laboratory
frame:
  \begin{equation}
  {\bf \tilde g}(t)={\bf O}^{-1}(\hat {\bf z}, \omega_L t)~{\bf g}~{\bf
O}(\hat {\bf z}, \omega_L t)~.
  \end{equation}
Here the matrix $O_{\alpha\beta}(\hat {\bf z}, \omega t)$ describes
the transformation from the laboratory frame into the rotating frame
- this is the rotation about the magnetic field axis $\hat {\bf z}$
by angle $\omega_L t$. Now we can find all the degenerate coherent
states of the Larmor precession applying the symmetry group $G$ to
the simplest equilibrium state of the given superfluid phase: ${\bf
A}={\bf O}^{-1}{\bf R}^{(1)} {\bf O}{\bf A}^{(0)}({\bf
R}^{(2)})^{-1}$, where ${\bf R}^{(1)}$ is the arbitrary matrix
describing spin rotations in the precessing frame and ${\bf
R}^{(2)}$ is another arbitrary matrix which describes the orbital
rotations in the laboratory frame. In case of $^3$He-B this state
corresponds to the state of Cooper pairs with $L=S=1$ and the total
angular momentum $J=0$ \cite{VollhardtWolfle}:
\begin{equation}
A^{(0)}_{\alpha i}=\Delta_B~\delta_{\alpha i}~~.
\end{equation}
  The action of elements of the group $G$ on this stationary state leads
to the following general precession of $^3$He-B with the Larmor
frequency (if the spin-orbit interaction is neglected):
\begin{equation}
A_{\alpha i}(t)=\Delta R_{\alpha i}(t)~~, \label{LarmorPrecession1}
  \end{equation}
   \begin{equation}
R_{\alpha i}(t)=O_{\alpha \beta}(\hat {\bf z}, -\omega
t)R^{(1)}_{\beta \gamma}O_{\gamma \mu}(\hat {\bf z}, \omega
t)(R^{(2)})^{-1}_{\mu i}~~ . \label{LarmorPrecession2}
  \end{equation}
The matrix ${\bf R}^{(1)}$ determines the direction of spin density
in the precessing frame:
\begin{equation}
  S_\alpha = \chi R^{(1)}_{\alpha\beta} H_{\beta}~~,
  \end{equation}
where $\chi$ is the spin susceptibility of $^3$He-B. This
corresponds to the precession of spin with the tipping angle
$\cos\beta_1=R^{(1)}_{zz}$. The matrix ${\bf R}^{(2)}$ determines
the direction of orbital momentum density in the laboratory frame:
\begin{equation}
   L_i =-R_{\alpha i}(t)S_{\alpha }(t)=-\chi R^{(2)}_{i\alpha }
H_{\alpha}~~,
  \end{equation}
with the tipping angle $\cos\beta_2=R^{(2)}_{zz}$.

\subsection{Spin-orbit interaction as perturbation}

The spin-orbit interaction in $^3$He-B is
  \begin{equation}
F_D={2\chi\Omega_L^2  \over 15}\Big[{\rm Tr}{\bf R}(t)-{1\over
2}\Big]^2 ={8\chi \Omega_L^2 \over 15}\Big[\cos\theta(t) +{1\over
4}\Big]^2, \label{SO}
  \end{equation}
where $\Omega_L$ is the so called Leggett frequency -- the frequency
of the longitudinal NMR; $\theta$ is the angle of rotation in the
parametrization of the matrix $R_{\alpha i}$ in terms of the angle
and axis of rotation \cite{VollhardtWolfle}; we use the system of
units in which the gyromagnetic ratio ${\gamma}$ for the $^3$He atom
is 1, hence the magnetic field and the frequency will have same
physical dimension.

  In the general state of the
Larmor precession (\ref{LarmorPrecession1}), the spin-orbit
interaction contains the time independent part and rapidly
oscillating terms with frequencies $\omega_L$, $2\omega_L$,
$3\omega_L$ and $4\omega_L$:
\begin{equation}
   F_D(\gamma)=F_0 + \sum_{n=1}^4F_n\cos(n\omega_Lt)~~.
\label{TotalFD}
  \end{equation}
The time-independent part -- the average over fast oscillations --
is
\begin{eqnarray}
  F_0={2\over 15}\chi\Omega_L^2[(sl-
{1\over 2}+{1\over 2}(1+s)(1+l)\cos\gamma)^2 +
   \nonumber\\
   {1\over 8}(1-s)^2(1-l)^2 +(1-s^2)(1-l^2)(1+\cos\gamma)] ~~.
   \label{gammaGeneral}
  \end{eqnarray}
Here $s=\cos\beta_1$ and $l=\cos\beta_2$ are $z$ projections of unit
vectors $\hat {\bf s}={\bf S}/S$ and $\hat {\bf l}=-{\bf L}/L$; and
$\gamma$ is another free parameter of the general precession.
Altogether the free precession is characterized by 5 independent
parameters coming from two matrices ${\bf R}^{(1)}$ and ${\bf
R}^{(2)}$ \cite{MisVol}: two angles of spin ${\bf S}$, two angles of
the orbital momentum ${\bf L}$, and the relative rotation of
matrices by angle $\gamma$. In the case of the non-precessing
magnetization, the $\gamma$-mode corresponds to the longitudinal NMR
mode.

  \section{Parametric instability of HPD to radiation of $\gamma$-mode}

  \subsection{Lagrangian for $\gamma$ mode}

  In the simplest description, the dynamics of the $\gamma$-mode is
determined by the following Lagrangian:
\begin{equation}
   {\cal L}= -\frac{1}{2}\chi \left(\dot \gamma^2
-c^2(\nabla\gamma)^2\right) + F_D(\gamma)~~.
   \label{Lagrangian}
  \end{equation}
Here we used the approximation of a single speed of spin waves $c$,
since the effect of the anisotropy of the spin wave velocity has
already been discussed in   Ref.~\cite{SF}. In the time-dependent part
of $F_D$ we only consider the first harmonic, i.e. according to
Eq.(\ref{resonance}) we discuss the parametric excitation of two
$\gamma$-modes with
$ck\approx \omega_L/2$. The amplitude of the first harmonic is:
\begin{eqnarray}
  F_1={4\over 15}\chi\Omega_L^2  \sin \beta_1 \sin \beta_2\cos(\gamma/2)
\times \nonumber\\
\left(2sl- 1 +\frac{(1-s)(1-l)}{2} +(1+s)(1+l)\cos\gamma\right) .
\label{FirstHarm}
  \end{eqnarray}
Further we assume that the system is in the minimum of the dipole
energy $F_0$ as a function of $\gamma$. The equilibrium value
$\gamma=\gamma_0$ is
\begin{equation}
\cos\gamma_0=- \frac{(2sl -1) +2(1-s)(1-l) } {(1+s)(1+l)}~~,
   \label{GammaMin}
  \end{equation}
which is valid if the right hand side of Eq. (\ref{GammaMin}) does
not exceed unity, i.e. when $s +l -5sl <2$.

For the discussion of Suhl instability we need the time-dependent
term which is quadratic in $\gamma-\gamma_0$. Then the Lagrangian
(\ref{Lagrangian}) which describes the parametric instability
towards decay of Larmor precession to two $\gamma$-modes with
$kc\approx \omega_L/2$ is (after the shift
$\gamma-\gamma_0\rightarrow \gamma$; neglecting $\Omega_L$ compared
to $\omega_L$; and neglecting the anisotropy of the spin-wave
velocity):
  \begin{equation}
   {\cal L}= \frac{1}{2}\chi \left(-\dot \gamma^2 +c^2(\nabla\gamma )^2+
a  \Omega_L^2 \gamma^2 \cos\omega_Lt \right)  ~,
\label{InstabilityLagrangian0}
  \end{equation}
where, if $s +l -5sl <2$, the parameter $a$ is
\begin{eqnarray}
  a=\frac{4}{15}  \sin \beta_1 \sin \beta_2\left[
\frac{3(s+l-sl)}{2(1+s)(1+l)}  \right]^{1/2} \times \nonumber
\\
\left[(1+s)(1+l) +2(2sl-1) +\frac{35}{8}(1-s)(1-l)\right]~.
\label{a1}
  \end{eqnarray}

\subsection{Parametric instability}

Let us rewrite the Lagrangian (\ref{InstabilityLagrangian0}) in
terms of Hamiltonian as function of creation and annihilation
operators $b_{\bf k}$ and $b_{\bf k}^*$:
  \begin{eqnarray}
  \gamma_{\bf k}&=&\frac {i \, (b_{\bf k}-b_{\bf
k}^*) }{\sqrt{2\chi \omega_s(k)}}\,, \quad \omega_s^2(k) = c^2k^2\,,
\\ p_{\bf k}&=&\chi\dot\gamma_{\bf k}= \sqrt{ \chi \omega_s(k)/2}
(b_{\bf k}+b_{\bf k}^*)~,
   \label{CreationAnnihilation}
\\
  {\cal H}&=&\sum_{\bf k} \omega_s(k) b_{\bf k}^*b_{\bf k}
   + \frac12 \sum_{\bf k}V(k)
 \left(e^{-i\omega_L t}b_{\bf
k}b_{-\bf k} + \mbox{c.c.}\right)\,,  \label{Hamiltonian}\\
\label{BLV}
 V(k)&=& V_{_{\rm BLV}}(k)\equiv a\Omega_L^2\Big /  \omega_s(k)\ .
\end{eqnarray}
Here we neglected $\Omega_L$ compared to $\omega_L$. The spectrum of
the excited mode is   $ b_{\bf k}(t)=\tilde b_{\bf
k}\exp(-i\omega_L/2 t +  \nu_{\bf k} t)$, where the instability
increment $\nu_{\bf k}$ is
  \begin{eqnarray}
\nu_{\bf k}=\sqrt{V(k)^2 -[\omega_s(k) -\omega_L/2]^2  }\ .
   \label{SpectrumExcited}
  \end{eqnarray}
Increment $\nu_{\bf k}$ reaches its maximum $\max \nu_k=V(k)$
for spin waves, satisfying condition~(\ref{resonance}) of the
parametric resonance, $\omega_s (k') =\omega_L/2$. These modes grow
exponentially:
 \begin{equation}
  b_{\bf k}(t) \propto  \exp (V(k') t)~ .
   \label{ExpGrowth}
  \end{equation}

At finite temperatures this growing is damped by dissipation, but at
low temperature the dissipation becomes small and catastrophic
relaxation occurs. We shall follow Ref. \cite{SF} and assume the
spin diffusion mechanism of dissipation. In this case the equation
for temperature $T_{\rm cat}$ below which the instability of the
homogeneous precession towards radiation of spin waves with
$\omega_s(k')=ck'=\omega_L/2$ starts to develop is:
\begin{equation}
  D(T_{\rm cat}) = 2V(k') c^2/\omega_L^2~.
   \label{Demp}
  \end{equation}
Here $D(T)$ is the spin diffusion coefficient, which depends on
temperature and decreases with decreasing $T$.

In our case
 \begin{equation}
 V_{_{\rm BLV}}(k')=2a\Omega_L^2/\omega_L~ .
   \label{ExpGrowthBLV}
  \end{equation}

\section{Two mechanism of Suhl instability}

The characteristic feature of spin-orbit interaction in the
isotropic $^3$He-B is that it is symmetric with respect to the
interchange of spin and orbital momenta \cite{MisVol}. This is the
reason why {\sl Eqs.} (\ref{gammaGeneral}), (\ref{FirstHarm}) and
(\ref{a1}) are symmetric under transformation
$\beta_1\leftrightarrow \beta_2$ (or $s\leftrightarrow l$). The
immediate consequence of this symmetry is that the Brinkman-Smith
(BS) precession ($\beta_1=\beta$, $\beta_2=0$) is equivalent to the
static state of $^3$He-B in magnetic field ($\beta_1=0$,
$\beta_2=\beta$). Since the static state is stable, the BS
precession cannot be destabilized by spin-orbit interaction, which
is also seen from Eq.(\ref{a1}): the interaction amplitude $V(k)=0$
if $\beta_2=0$ ($l=1$).

This $Z_2$ symmetry of the Larmor precession in $^3$He-B is violated
by the term in the gradient energy which is responsible for the
dependence of the spin-wave velocity on the direction of
propagation. This term, which we omitted in our consideration,
becomes time dependent even in the background of the BS mode. This
leads to the Sourovtsev-Fomin (SF) contribution $V_{_{\rm SF}}(k)$
to the interaction amplitude $V(k)$,   discussed by \cite{SF} for
$\bf L || \bf H$.  SF mechanism can be easily extended to the
general precession with arbitrary $\beta_2$. For that one must take
into account, that the axis of anisotropy of spin-wave propagation
is determined by the orbital vector ${\bf L}$. This gives the
following modification of $V_{_{\rm SF}}(k')$:
\begin{eqnarray}
V_{_{\rm SF}}(k') \Rightarrow \widetilde V_{_{\rm SF}}(k')
=\frac{\mu\,  \omega_L }{4}\frac{\sin\beta_1|1-2s|} {2s^2-2s +5}
|\sin~2\delta|\,, \label{SF}\\  \cos\delta\equiv \hat{\bf k}\cdot \hat{\bf
l}\ .
   \label{SFdelta}
  \end{eqnarray}
Here $\mu=1-c_\perp^2/c_\parallel^2$ is the anisotropy of the
spin-wave velocity, where $c_\parallel$ ($c_\perp$) is the velocity
of spin waves propagating along (perpendicular to) $\hat{\bf l}$. We
assume that both $\Omega_L^2/\omega_L^2$ and the spin-wave
anisotropy $\mu$ are small parameters. In general

\begin{equation}
V(k)=V_{_{\rm BLV}}(k) + \widetilde V_{_{\rm SF}}(k)\ .
\label{gen}
\end{equation}

Comparing these two contributions to $V(k)$, causing the
catastrophic relaxation -- due to spin-orbit interaction,
$V(k)=V_{_{\rm BLV}}(k)$,   Eq.~(\ref{BLV}) and due to spin-wave
anisotropy in Eq.~(\ref{SF}) -- one can see  that  $ \widetilde
V_{_{\rm SF}}(k')>V_{_{\rm BLV}}(k') $   at high fields, when the
Larmor frequency $\omega_L$ essentially exceeds the Leggett
frequency $\Omega_L$. However, in the typical experiments \cite{HP1,
HP2} the spin-orbit amplitude, $ V_{_{\rm BLV}}(k)$,   appears to be
more important. There are two reasons for that: the field is not
sufficiently high ($\Omega_L/\omega_L\sim 0.3-0.5$); the texture of
the orbital $\hat{\bf l}$ is typically formed due to boundary
conditions at the walls of container: $\hat{\bf l}$ must be oriented
along the normal to the wall. The 1D simulation \cite{BG} of the
texture of the $\hat{\bf s}$ and $\hat{\bf l}$ fields (angles
$\beta_1$ and $\beta_2$) under conditions of experiments but in the
parallel-plate geometry is shown in Fig.1. The main part of the
deviation from the BS mode develops in about 1 mm size region near
the walls, which means that in the cylindrical cell the texture
should be very similar. Thus the real mode of precession in the
finite vessel is very different from the ideal BS configuration,
which is static in the precessing frame. On the contrary, the angle
$\theta$ of the order parameter $R_{\alpha i}$ is oscillating as
shown in the upper part of the Figure. These oscillations lead to
 parametric
instability at low temperature.

\begin{figure}
 \includegraphics[height=7cm]{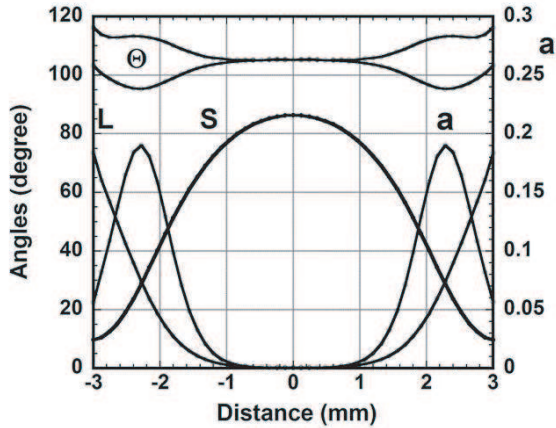}
\caption{\label{figure} The texture in the precessing state in the
cell between two parallel walls at $x= 0$ and $x=6$ mm. Left scale
shows angles $\beta_1$, $\beta_2$ and $\theta$, while the right
scale shows the local value of the parameter $a$ in Eq.(\ref{a1}).}
\end{figure}

Figure 1 also shows the parameter $a(x)$ as function of the
coordinate $x$. Since the diffusion damping of spin waves occurs in
the whole volume of the cell, the increment $V(k')$ in
Eq.(\ref{ExpGrowthBLV}) must be averaged over the cylindrical cell:
$V(k')=2\bar a
\Omega_L^2/\omega_L$, with $\bar a$= 0.099 for our texture. This
leads to the critical value of spin diffusion at which the
catastrophic relaxation must occur
\begin{equation}
  D(T_{\rm cat}) = 4 \bar a \frac{\Omega_L^2
c_\parallel^2}{\omega_L^3}=0.085   {\rm cm}^2/{\rm s} \label{Demp1}
\end{equation} For $\Omega_L=$244 kHz and $\omega_L=$ 460kHz this
value corresponds to the temperature of the catastrophic relaxation
$T_{\rm cat} =0.5 T_c$, according to \cite{BE}, in a good agreement
with the experimental value $T_{\rm cat} = 0.47 T_c$
(\cite{HP1,HP2}).

In Ref. \cite{SF} the critical diffusion for optimal configuration
for spin wave velocity anisotropy was estimated as about 0.03
$cm^2/c$. Because the texture of the $\hat {\bf l}$-vector
influences $\widetilde V_{_{\rm SF}}(k')$ in Eq.(\ref{SF}), the increment
must be even smaller. This indicates that the spin-orbit mechanism of
catastrophic relaxation is dominating under the conditions of the
experiments \cite{HP1, HP2},  $ V_{_{\rm BLV}}(k') >
\widetilde V_{_{\rm SF}}(k')$.

To verify that it is really so, let us analyze the Grenoble
experiment \cite{B}, in which the texture was destroyed by the RF
pulse so that ${\bf L}$ was parallel to ${\bf H}$ even in the
surface layer. Under these conditions the catastrophic relaxation
was observed at lower temperature of about 0.4 T$_c$. In this
configuration one has $\beta_2=0$, the spin-orbit term, $V_{_{\rm
BLV}}(k)$, is switched off, the spin-wave anisotropy of Ref.
\cite{SF}, leading to $V_{_{\rm SF}}(k)$,  becomes the main source
of instability, and indeed this $T_{\rm cat}$ well corresponds to
the critical spin diffusion $D(T_{\rm cat})$ about 0.03 $cm^2/c$,
calculated in Ref. \cite{SF}. When the texture is restored, then if
the spin-wave anisotropy is the only mechanism of the catastrophic
relaxation, the texture should lead to decrease of $V(k)$ and
thus to decreasing $T_{\rm cat}$. Instead, $T_{\rm cat}$ increases
demonstrating that when the texture appears another mechanism of
instability emerges as we discussed here.

Two contributions, Eq.~(\ref{gen}),  to the increment of parametric
instability~(\ref{SpectrumExcited}) can be also compared using
results of experiments made in Cornell \cite{Lee}. In these
experiments $T_{\rm cat}$ was measured for 31 bar at different
$\omega_L$. They found that $T_{\rm cat}$ decreases from 0.39 T$_c$
at $\omega_L=600$ kHz to 0.24 T$_c$ at $\omega_L= 3$ MHz. This is
well described by the spin-orbit mechanism  of Suhl instability,
which becomes weaker at higher field according to Eq.
(\ref{ExpGrowth}). On the contrary, if the spin-wave anisotropy is
the dominating mechanism, $T_{\rm cat}$ would not depend on
$\omega_L$ (there are  processes in  the spin-wave anisotropy
mechanism whose increment is proportional to $\Omega_L^2/\omega_L^2$
Ref.\cite{SF} as in the case of the spin-orbit mechanism, but they
are relatively small, since contain the product of two small
parameters, $\mu$ and $\Omega_L^2/\omega_L^2$) .

\section{Conclusion}

We   calculated in a unified way two contributions   $
V_{_{\rm BLV}}(k)$ and $\widetilde V_{_{\rm SF}}(k)$ to the
increment of the parametric instability of the Larmor precession in
superfluid $^3$He-B for any angle between $\bf L$ and $\bf H$.
The  term   $V_{_{\rm SF}}(k)$,   previously  discussed in Ref.
\cite{SF} for simple geometry with  ${\bf L} || \bf H $, originates
from the anisotropy of spin-wave velocity. We modified this
contribution, $ V_{_{\rm SF}}(k) \Rightarrow\widetilde V_{_{\rm
SF}}(k) $  for the case of the texture of the order parameter, which
occurs in real experiments. In addition we found another term,  $
V_{_{\rm BLV}}(k)$, which originates from the spin-orbit
interaction. This second mechanism only takes place when the orbital
momentum ${\bf L}$ deviates from its symmetric orientation along the
magnetic field, in particular in the presence of texture. Our
analytical result for the onset of
the parametric
instability due to this mechanism is in a good quantitative
agreement with experimental results. In particular it gives the
correct dependence of the catastrophic relaxation temperature on
magnetic field. The spin-wave anisotropy mechanism must dominate
either when the texture is destroyed or in the limit of strong
magnetic field.

This work was done as the result of collaboration in the framework
of the ESF Program COSLAB, the Large Scale Installation Program
ULTI of the European Union (contract number: RITA-CT-2003-505313)
and the project ULTIMA of the "Agense National de Recherche",
France.  The work was also supported in part by the Russian
Foundation for Fundamental Research and the US-Israel Binational
Science Foundation.

\end{document}